\documentclass[sigconf, screen, nonacm]{acmart}

\AtBeginDocument{%
  \providecommand\BibTeX{{%
    \normalfont B\kern-0.5em{\scshape i\kern-0.25em b}\kern-0.8em\TeX}}}

\acmConference[Preprint]{Preprint}{2020}{USA}
\acmBooktitle{2020}

\usepackage{adjustbox}

\begin{document}


\title[CheXpedition]{CheXpedition: Investigating Generalization Challenges for Translation of Chest X-Ray Algorithms to the Clinical Setting}

\author{Pranav Rajpurkar}
\authornote{Both authors contributed equally to this research.}
\email{pranavsr@cs.stanford.edu}
\orcid{1234-5678-9012}
\affiliation{%
  \institution{Stanford University}
}

\author{Anirudh Joshi}
\email{anirudhjoshi@stanford.edu}
\authornotemark[1]
\affiliation{%
  \institution{Stanford University}
  }

\author{Anuj Pareek}
\email{anujpare@stanford.edu}
\orcid{0000-0002-1526-3685}
\affiliation{%
  \institution{Stanford University}
  }

\author{Phil Chen}
\author{Amirhossein Kiani}
\author{Jeremy Irvin}
\affiliation{%
  \institution{Stanford University}
  }

\author{Andrew Y. Ng}
\affiliation{%
  \institution{Stanford University}
  }

\author{Matthew P. Lungren}
\affiliation{%
  \institution{Stanford University}
  }

\renewcommand{\shortauthors}{Rajpurkar et al.}

\begin{abstract}
Although there have been several recent advances in the application of deep learning algorithms to chest x-ray interpretation, we identify three major challenges for the translation of chest x-ray algorithms to the clinical setting. We examine the performance of the top 10 performing models on the CheXpert challenge leaderboard on three tasks: (1) TB detection, (2) pathology detection on photos of chest x-rays, and (3) pathology detection on data from an external institution. First, we find that the top 10 chest x-ray models on the CheXpert competition achieve an average AUC of 0.851 on the task of detecting TB on two public TB datasets without fine-tuning or including the TB labels in training data. Second, we find that the average performance of the models on photos of x-rays (AUC = 0.916) is similar to their performance on the original chest x-ray images (AUC = 0.924). Third, we find that the models tested on an external dataset either perform comparably to or exceed the average performance of radiologists. We believe that our investigation will inform rapid translation of deep learning algorithms to safe and effective clinical decision support tools that can be validated prospectively with large impact studies and clinical trials.
\end{abstract}



\begin{teaserfigure}
  \includegraphics[width=\textwidth]{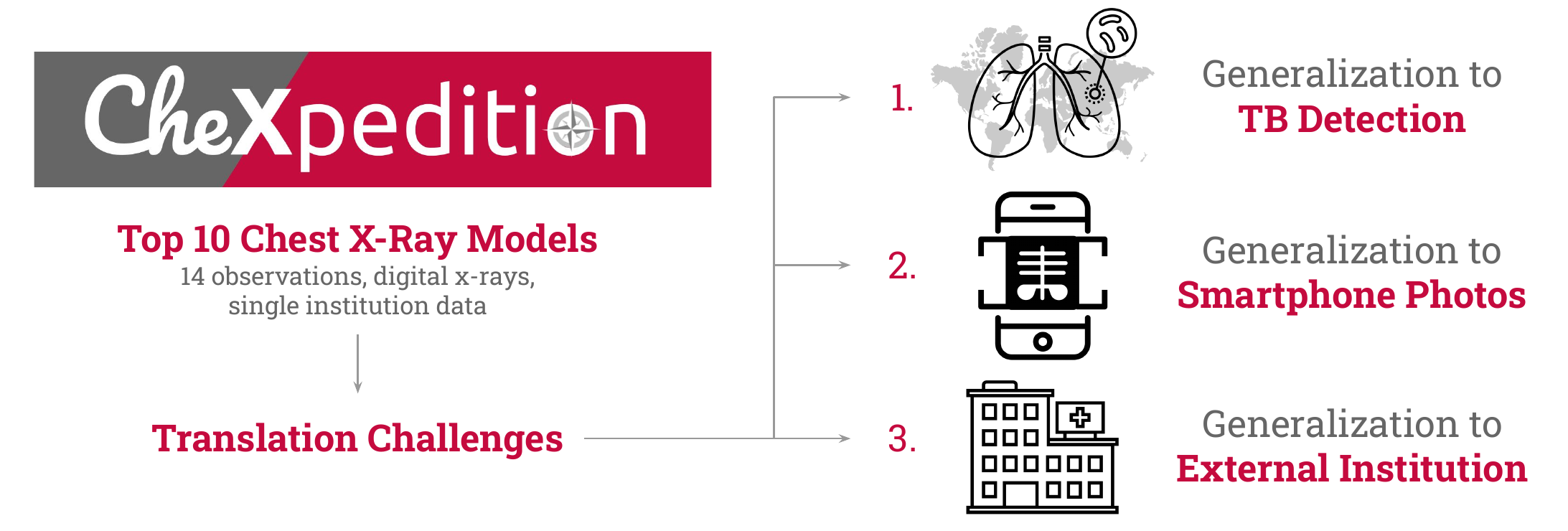}
  \caption{We examine the performance of the top 10 available models on the CheXpert competition leaderboard on three tasks: (1) TB detection, (2) pathology detection on photos of chest x-rays, and (3) pathology detection on data from an external institution}
  \label{fig:teaser}
\end{teaserfigure}

\maketitle

\section{Introduction}
There have been several recent advances in the application of deep learning algorithms to chest x-ray interpretation at a high level of performance \citep{rajpurkar_deep_2018, singh_deep_2018, nam2018development}. Although these advancements have led many to suggest a near-term potential of these algorithms to provide accurate chest x-ray interpretation and increase access to radiology expertise, a few major challenges remain to their translation to the clinical setting.

There remain major challenges for the translation of chest x-ray algorithms to the clinical setting. First, the performance of deep learning chest x-ray algorithms, trained with mainly US-based chest x-ray datasets, on endemic and globally relevant diseases not commonly found in the US, such as tuberculosis (TB) is unknown \citep{qin_computer-aided_2018, qin_using_2019}. Second, most chest x-ray algorithms have been developed and validated on digital x-rays, while the vast majority of the world relies on film for X-ray interpretation, a barrier that denies these populations from the advancements of automated interpretation \citep{films}. In order to apply an interim digital solution, digital photographs of films for storage, interpretation, and consultation can be performed as a "workaround" \cite{handelman_rogers_babiker_lee_mcmonagle_2018}. Third, chest x-ray algorithms which are developed using the data from one institution have not shown sustained performance when externally validated in application data from a different unrelated institution, and instead, these models have been criticized as vulnerable to bias and non-medically relevant cues \citep{zech_variable_2018}. We believe that tackling each of these challenges will serve to inform improved translation of deep learning algorithms into safe and effective clinical decision support tools that can be validated prospectively with large impact studies and clinical trials.

The purpose of this work is to systematically address the aforementioned translation challenges for chest x-ray models. We validate the performance of chest x-ray models on the tasks of (1) TB detection (2) pathology detection on digital photographs of chest x-rays, and (3) pathology detection on chest x-rays from a separate institution. Rather than choosing one model architecture or approach, we evaluate performance under each of the conditions using the top 10 performing models on the CheXpert challenge, a large public competition for chest x-ray analysis \citep{irvin_chexpert:_2019}. 

In this work we report performance metrics for the generalizability of existing chest x-ray models on the three aforementioned tasks. First, we find that the top 10 chest x-ray models on the CheXpert competition without fine-tuning or including the TB labels in training data, achieve an average AUC of 0.851 on the task of detecting TB on two public TB datasets, competitive with previously published approaches that trained and tested their models specifically on these same TB datasets \citep{qin_computer-aided_2018, qin_using_2019}. Second, we find that the average performance of the models on photos of x-rays (AUC = 0.916) is similar to their performance on the original chest x-ray images (AUC = 0.924). Third, we find that the models tested on an external dataset either perform comparably to or exceed the average performance of radiologists.

\begin{figure}[ht!]
    \includegraphics[width=\columnwidth]{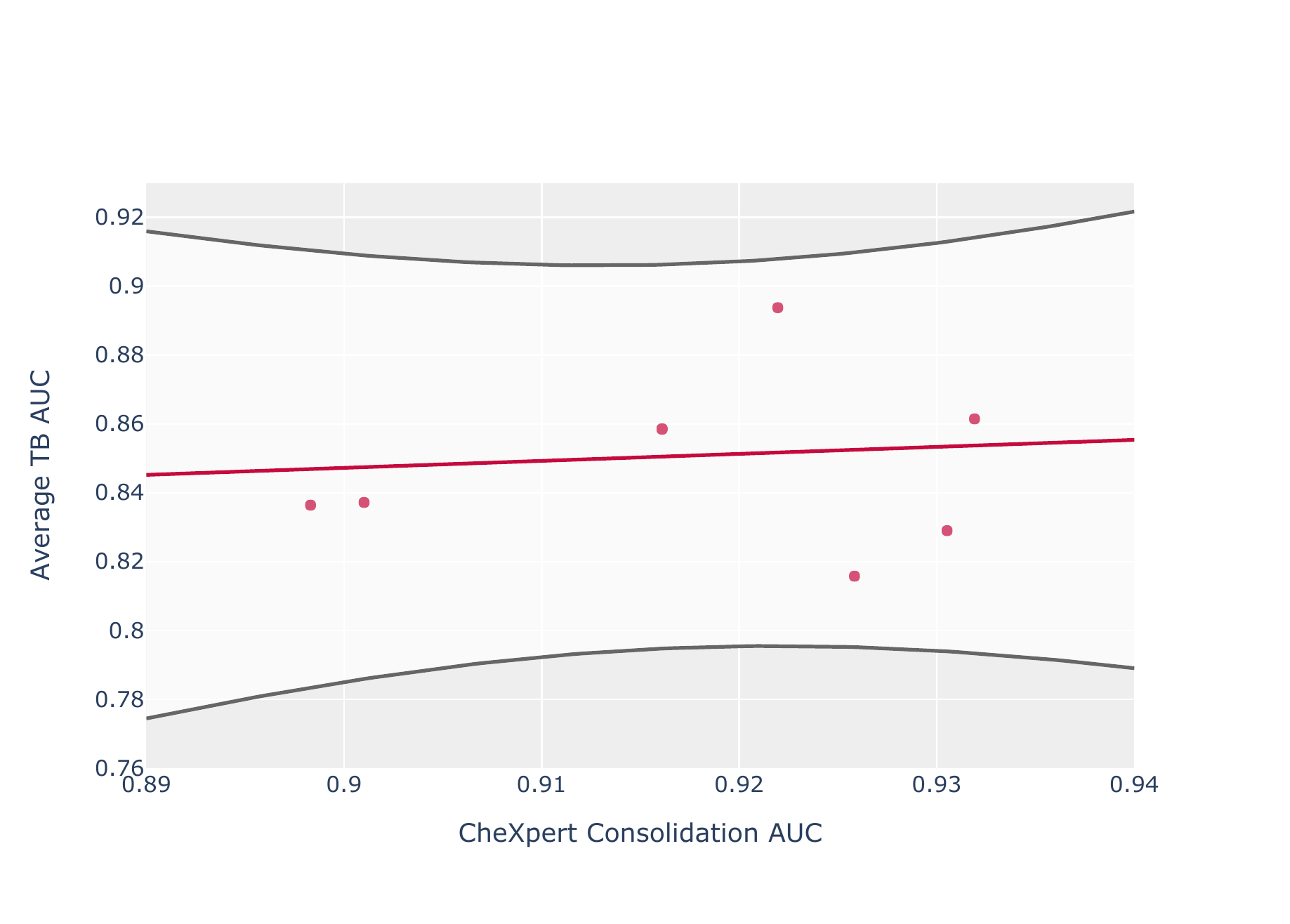}
    \includegraphics[width=\columnwidth]{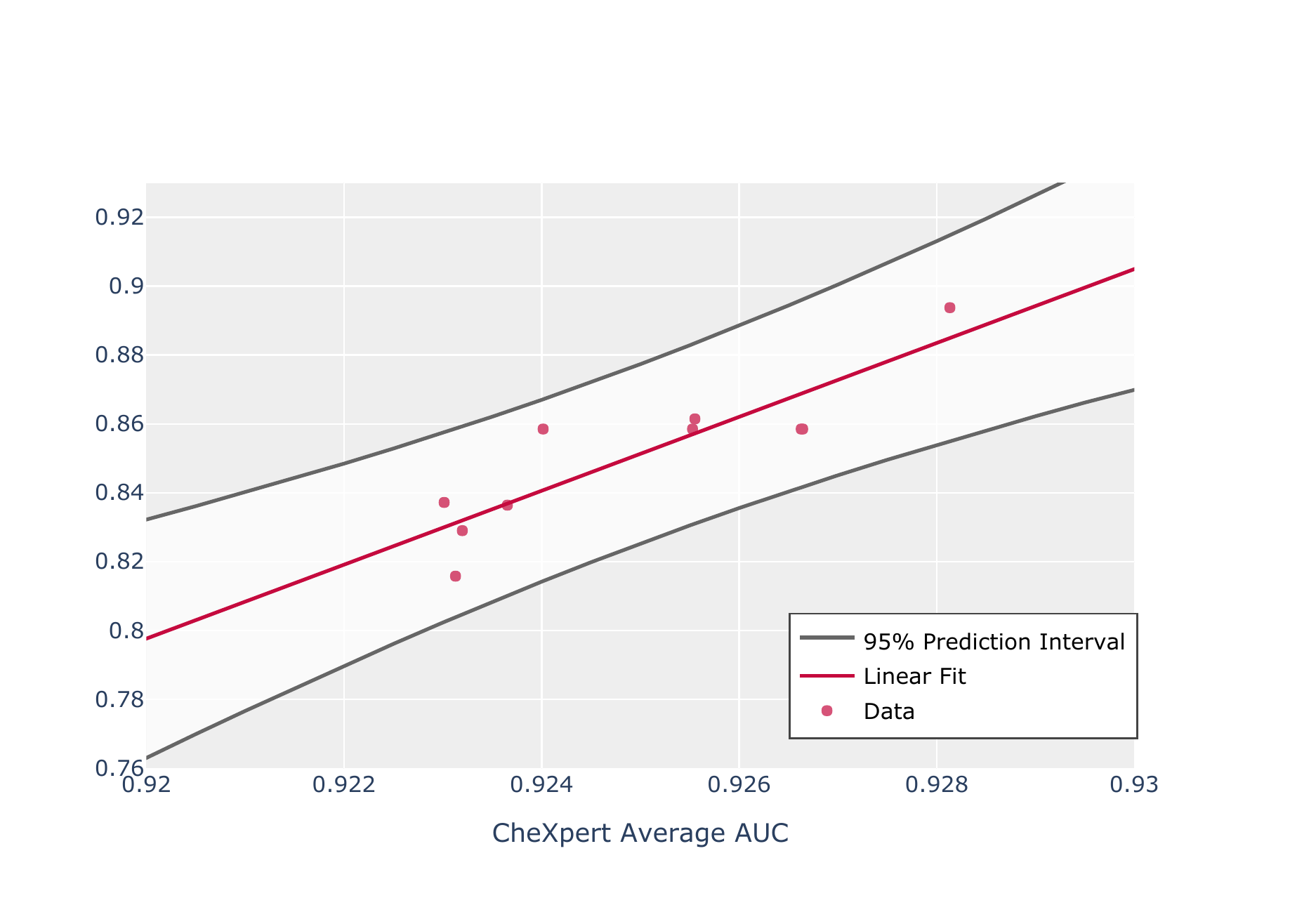}
    \caption{The average AUC of the models on all five CheXpert tasks was a stronger predictor (R2=0.78) of TB performance than the performance of 
    on the consolidation task in CheXpert (R2=0.011).}
    \label{fig:tb_auc}
\end{figure}
\section{Experimental Setup}

\subsection*{Top Models on CheXpert Leaderboard}
We investigated the generalization performance of the top 10 models on the CheXpert \citep{irvin_chexpert:_2019} competition leaderboard. CheXpert is a competition for automated chest x-ray interpretation that has been running from January 2019 featuring a strong radiologist-labeled reference standard.
As of November, 2019, there were 94 models that had been submitted to the CheXpert leaderboard from both academic and industry teams. The top 10 available models on the CheXpert competition leaderboard as of November 2019 were selected. All of the selected models were ensembles with the number of models in the ensemble ranging from 8 to 32; the majority of these models featured Densely Connected Convolutional Networks \cite{huang_densely_2017} as part of their ensemble.

\subsection*{Running Models on New Test Sets}
CheXpert is a unique competition in that it uses a hidden test set for official evaluation of models. Teams submit their executable code, which is then run on a test set that is not publicly readable. Such a setup preserves the integrity of the test results. Models can be rerun on new test sets to evaluate the ability of the model to generalize to new domains.

We make use of the CodaLab platform to re-run these chest x-ray models on the new test sets. CodaLab is an online platform for collaborative and reproducible computational research. The system exposes a simple command-line interface using which one can upload code and data and subsequently submit jobs to run them. Once a team has submitted their model on CodaLab and successfully inferred on the hidden CheXpert test set, they get added to the leaderboard. We reproduced the runs of the top 10 teams using their model checkpoints and inference scripts by substituting the hidden CheXpert test set for the other datasets used in this study.

\subsection*{Evaluation Metrics}
Our primary evaluation metric is the area under the receiver operating characteristic curve (AUC). We report the average AUC of the top 10 models on the new test sets, averaged over the available tasks. Additionally, in experiments comparing the models to board-certified radiologists, we compute the average sensitivities of all models per task thresholded at the specificities of the radiologists at each task.  The sensitivities of the average of the radiologists are compared to the sensitivities of the models per task.

\subsection*{Saliency Maps}
Gradient-weighted class activation maps (CAMs) \cite{gradcam, cam} were used to highlight regions with the greatest influence on a model’s decision. For a given x-ray, the CAM was produced for every class by taking the weighted average across
the final convolutional feature map, with weights determined by the linear layer. The CAM was then scaled according to the output probability, so that more confident predictions appeared brighter. Finally, the map was
upsampled to the input image resolution, and overlaid onto the input image. The Stanford baseline model on the CheXpert leaderboard was used as the model of choice to generate the CAMs.

\begin{figure*}[ht!]
\centering
\includegraphics[width=0.7\textwidth]{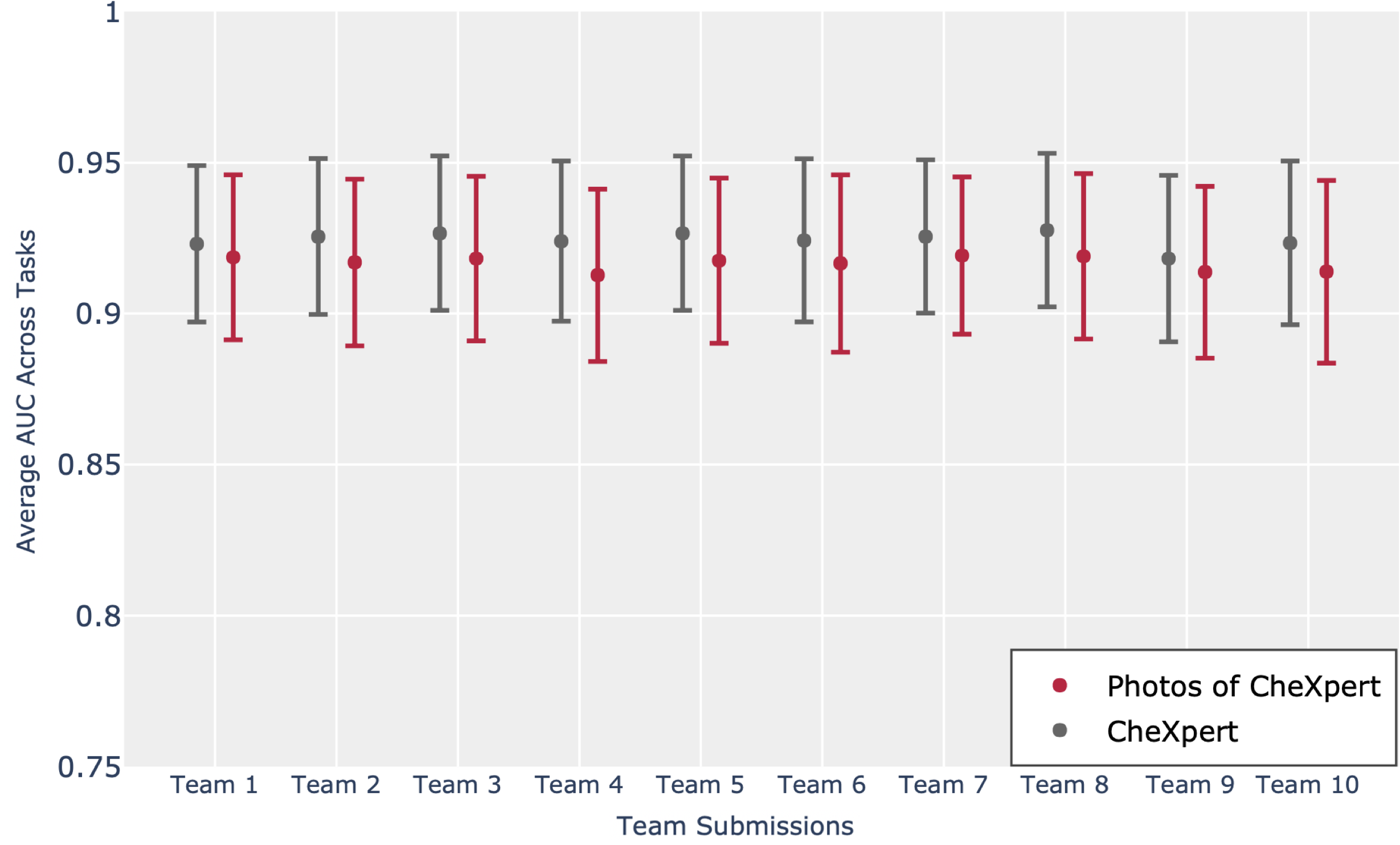}
\caption{The average performance of the models on photos of x-rays (AUC =  0.916) is similar to their performance on the original x-ray images (AUC = 0.924).}
\label{fig:photos}
\end{figure*}

\begin{figure}[ht!]
\centering
\includegraphics[width=\columnwidth]{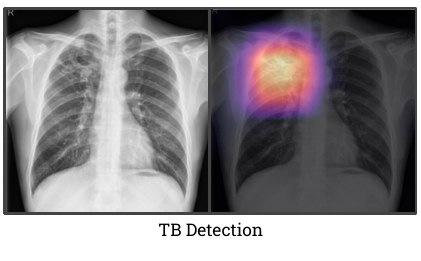}
\caption{The CAM for the TB detection task on the consolidation class correctly highlights the region of the image consisting with TB.}
\label{fig:photos_cam}
\end{figure}

\section{TB Detection}
\subsection*{Task}
We evaluated the models on the task of detecting tuberculosis (TB). TB is the leading cause of death from a single infectious disease agent and the leading cause of death for people living with human immunodeficiency virus (HIV) infection \cite{tb_stats}. Currently, chest x-ray models trained using large datasets from American institutions \cite{irvin_chexpert:_2019, johnson_mimic-cxr-jpg_2019, wang2017chestx} do not include TB as one of the labeled pathologies because the pathology is not prevalent in their settings. However, the application of chest x-ray models to the global setting requires their high performance on this globally relevant task.
We hypothesized that we could use existing models trained on the CheXpert dataset to detect TB without any fine-tuning on the TB task or the TB datasets. Because consolidation is one of the most common chest x-ray manifestations of pulmonary TB, we considered the use of the consolidation label as a proxy for the task of detecting TB.

\subsection*{Datasets}
We tested the performance of the models on two datasets: the Shenzhen and Montgomery datasets released by the NIH \cite{tb_data}. The Shenzhen dataset was collected in the Shenzhen No.3 Hospital, China. Of the 662 x-rays in the dataset, 326 are normal and 336 are abnormal with manifestations of TB; 34 cases are pediatric cases (defined as age $<$ 18 years). The Montgomery set was collected by the Department of Health and Human Services in Montgomery County, USA. Of the 138 x-rays in the dataset, 80 are normal and 58 are abnormal; 17 cases are pediatric cases.

\subsection*{Results} We evaluated the performance of the models using their probability on the consolidation label as the predicted score for TB on an x-ray (see Figure \ref{fig:tb_auc}). The average AUC of the models on the TB test sets ranged from 0.815 to 0.893 with an average of 0.851. 

\subsection*{Analysis} We analyzed the strength of the relationship between the performance of the models on the source tasks and the target TB dataset. We ran a linear regression to predict the average AUC of the models on the TB datasets using (1) the average AUC of the models on the consolidation task in CheXpert, and (2) the average AUC across all 5 competition tasks in CheXpert.

We found that the strength of relationship was smaller for the AUC on consolidation in CheXpert ($R^2 = 0.011$) compared to the average AUC on all five CheXpert tasks ($R^2 = 0.78$).

\subsection*{Discussion}
There have been a number of studies developing models for TB detection. \citet{hwang_novel_2016} tested on the Shenzhen TB dataset without training on the data, but their models were explicitly trained on the TB task, and achieved an AUC of 0.884 on the Shenzhen dataset. \citet{pasa_efficient_2019} reported AUCs of 0.811 on Montgomery and 0.900 on the Shenzhen dataset when their model was trained on a combination of the two datasets and additional data. Similarly, \citet{vajda2018feature} reported AUCs of 0.870 on Montgomery and 0.990 on the Shenzhen dataset after training on the same two datasets. Finally, \citet{lakhani_deep_2017} trained on a combination of four different TB datasets, and achieved an AUC of 0.990 on their test set with their ensemble model.

In our study, we found that the average AUC of the models on the TB test sets (average AUC of 0.851) without exposure to TB datasets was competitive to that of models that had been directly trained on these datasets for the task of tuberculosis detection. We also found that the average performance of a model across tasks was a stronger predictor of performance on the tuberculosis dataset as compared to the performance of the model on any of the individual tasks. This suggests that training models to perform well across tasks may allow them to perform better on unseen images than models that optimize for a single task. A possible reason for this finding may be that the shared representations learnt by optimizing for multitask performance are exploited for better performance on different data distributions \cite{multitask}.

\begin{figure*}[ht!]
\centering
\includegraphics[width=0.6\textwidth]{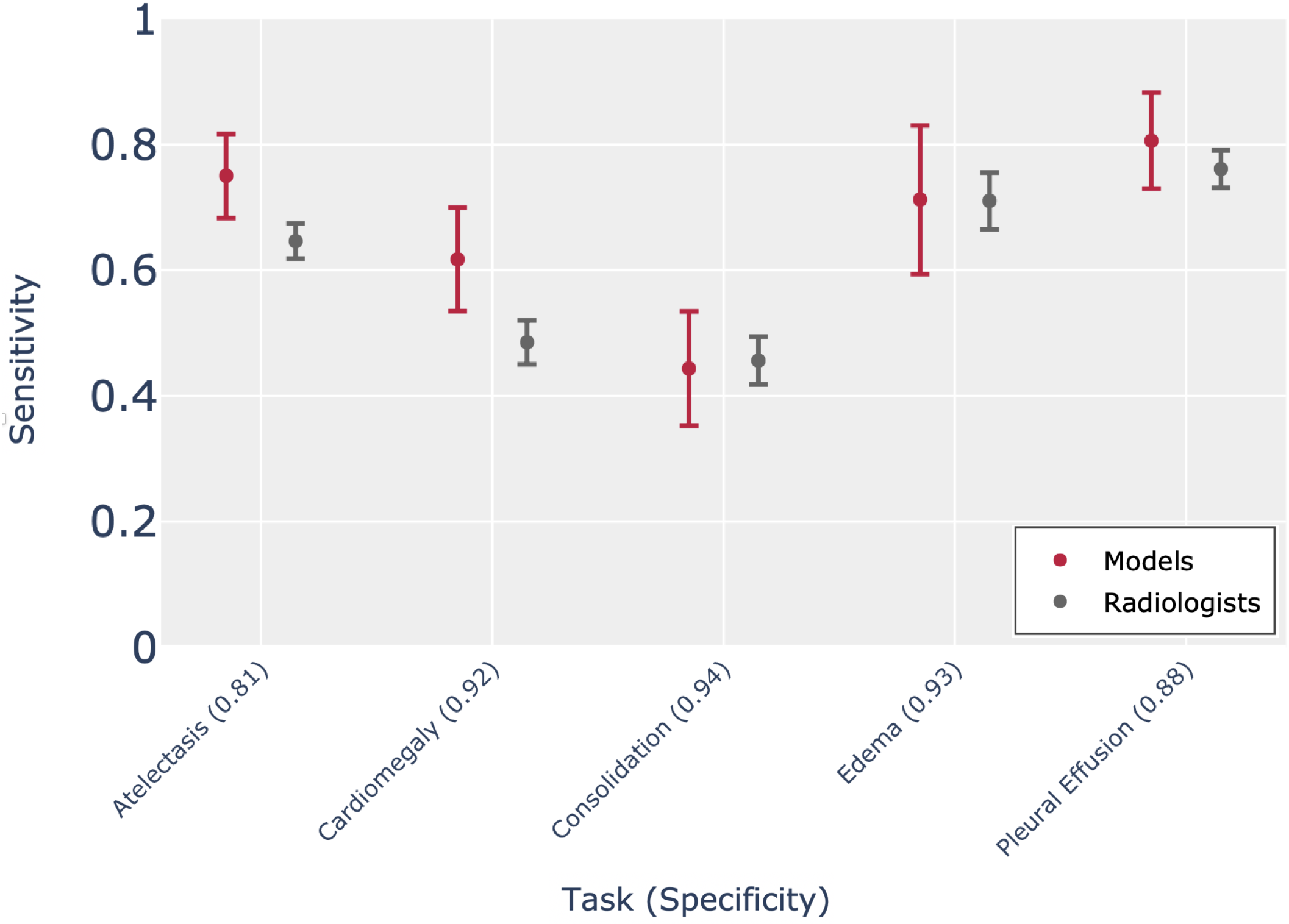}
\caption{The average performance of models tested on the external dataset is either comparable to or exceeds the average performance of radiologists on all 5 tasks.}
\label{fig:nih}
\end{figure*}
\begin{figure}[ht!]
\centering
\includegraphics[width=0.5\textwidth]{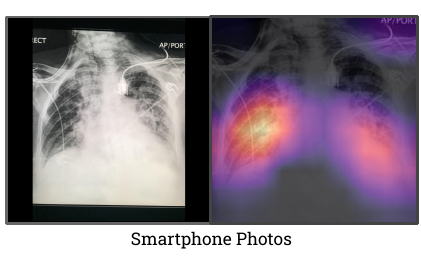}
\caption{The CAM for the photo of the portable frontal radiograph of the chest demonstrates cardiomegaly and bilateral mid and lower lung interstitial predominant opacities consistent with pulmonary edema.}
\label{fig:nih_cam}
\end{figure}

\section{Smartphone Photos}
\subsection*{Task}
We evaluated the models on the task of detecting pathologies on smartphone photos of chest x-rays. While most deep learning models are trained on digital x-rays, scaled deployment demands a solution that can navigate an endless array of medical imaging / IT infrastructures. An appealing solution to scaled deployment is to leverage the ubiquity of smartphones: clinicians and radiologists in parts of the world take smartphone photos of medical imaging studies to share with other experts or clinicians using messaging services like WhatsApp \cite{handelman_rogers_babiker_lee_mcmonagle_2018}. While using photos of chest x-rays to input into chest-xray algorithms could enable any physician with a smartphone to get instant AI algorithm assistance, the performance of chest x-ray algorithms on photos of chest x-rays has not been thoroughly investigated. Outside chest x-ray classification, deep learning algorithms for image classification have been shown to attain lower performance on photos of images than on the images themselves \cite{adversarial_examples}. We conducted an experiment to determine whether existing chest x-ray models could generalize well to photos of chest x-rays.

\subsection*{Datasets}
We generated a dataset of photos of the CheXpert test set, consisting of studies from 500 patients. Chest X-rays from each test study were displayed on a non-diagnostic computer monitor. Photos of the monitor were taken with an Apple iPhone 7 by a physician. The physician was instructed to keep the mobile camera stable and center the lung fields in the camera view. A time-restriction of 5 seconds per image was imposed to simulate a busy healthcare environment. Subsequent inspection of photos showed that they were taken with slightly varying angles; some photos included artefacts such as Moiré patterns and subtle screen-glares. Photos were labeled using the ground truth for the corresponding digital x-ray image.

\subsection*{Results}
The models achieved a mean AUC of 0.916 on photos of the chexpert test set, compared with an AUC of 0.924 on the original chexpert test set. All of the models had mean AUCs higher than 0.9, and were within 0.01 AUC of their performance on the original images. The average AUCs of each of the top 10 models across the 5 CheXpert competition tasks are detailed in Figure \ref{fig:photos}.

\subsection*{Discussion}
Several studies have highlighted the importance of generalizability of computer vision models with noise in images \citep{hendrycks_benchmarking_2019}. \citet{dodge_study_2017} demonstrated that deep neural networks perform poorly compared to humans on image classification on distorted images. \citet{schmidt_adversarially_2018, geirhos_imagenet-trained_2019} have found that convolutional neural networks trained on specific image corruptions did not generalize, and the error patterns of network and human predictions were not similar on noisy and elastically deformed images. 

In our study, the dataset we generated for this experiment allows for the direct comparison of the effect of photos against the source images on model performance, addressing a key deployment and generalization challenge. We found that the performance across top teams on photos of chest x-rays was comparable to their performance on the original x-rays. Figure \ref{fig:nih_cam} demonstrates that the model is able to detect the location of the pathology on a characteristic example where the distortion generated by taking photos of the x-rays did not affect the ability of the model to identify clinically relevant information in the x-rays.
\begin{table*}[ht!]
  \centering
  \begin{tabular}{|p{0.4\textwidth}|p{0.4\textwidth}|p{0.1\textwidth}|}
    \hline
    Example & Category & Count\\
    \hline
    \begin{minipage}{0.4\textwidth}
      \includegraphics[width=\linewidth]{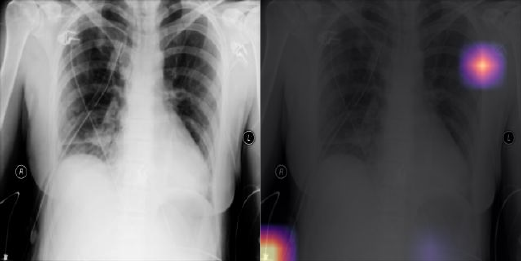}
    \end{minipage}
    &
    \textbf{Failed to correctly localize (False Negative)}. CAMs failed to localize the actual consolidation. Typically, the consolidation was smaller or less opaque than average; in some cases, the CAMs highlighted a feature that was visually similar but unrelated to consolidation.
    & 
    36 (44.44\%)
    \\
    \hline
    \begin{minipage}{0.4\textwidth}
      \includegraphics[width=\linewidth]{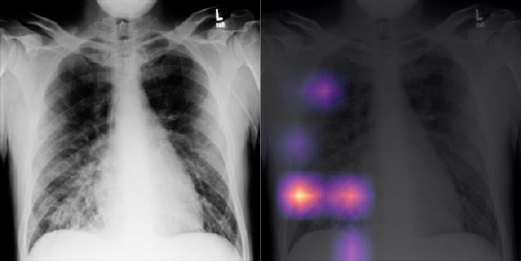}
    \end{minipage}
    &
    \textbf{Failed to confidently detect (False Negative)}. CAMs accurately localized the consolidation, but wasn't confident enough to make a positive diagnosis. This was found to occur when the consolidation was overlapping with other diseases (such as severe pulmonary edema) or anatomical structures.
    & 
    29 (35.80\%)
    \\
    \hline
    \begin{minipage}{0.4\textwidth}
      \includegraphics[width=\linewidth]{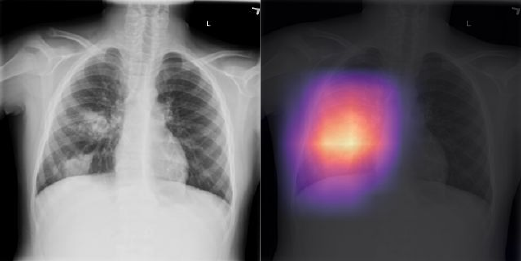}
    \end{minipage}
    &
    \textbf{Mistaken for mimicking feature (False Positive)}. CAMs detected a visual feature which mimics consolidation and made a false positive diagnosis. This was often the case in the presence of severe pulmonary edema, and cases with other pulmonary opacities such as fibrosis, scarring and lung lesion.
    & 
    13 (16.05\%)
    \\
    \hline
    \begin{minipage}{0.4\textwidth}
      \includegraphics[width=\linewidth]{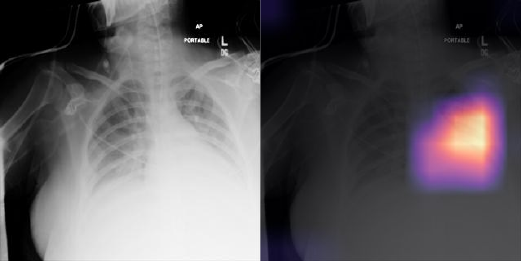}
    \end{minipage}
    &
    \textbf{Mistaken for non-mimicking feature (False Positive)}. The x-ray contains enlarged cardiac contours and bilateral mid and lower lung interstitial predominant opacities consistent with cardiomegaly and pulmonary edema. CAMs highlighted an area of the cardiac border and chest wall which bear no apparent visual resemblance to consolidations. 
    & 
    3 (3.70\%)
    \\
    \hline
  \end{tabular}
  \caption{Qualitative CAM Analysis of CheXpert model mistakes (73 images out of 420 images) on the consolidation task on external institution test data.}\label{tbl:cam_analysis}
\end{table*}

\section{External Institution} 

\subsection*{Task}
We evaluated the performance of the top 10 CheXpert models on a dataset from an external institution. Chest x-ray algorithms which are developed using the data from one institution have not shown sustained performance when externally validated in application data from a different unrelated institution and have been criticized as vulnerable to bias and non-medically relevant cues \citep{zech_variable_2018}. Furthermore, certain institutions may not allow access to patient data for privacy reasons. This makes it important for models trained on one institution's data to be generalizable to others without finetuning or retraining for wider deployment in the healthcare system.

\subsection*{Dataset}
We used a set of 420 frontal chest x-rays curated in the test set of \citet{rajpurkar_deep_2018}. These x-rays contained images from the ChestXray-14 dataset collected at the National Institutes of Health Clinical Center \cite{wang2017chestx}, sampled to contain at least 50 cases of each pathology according to the original labels provided in the dataset.

\subsection*{Results}
The models achieved an average performance of 0.897 AUC across the 5 CheXpert competition tasks on the test set from the external institution. On Atelectasis, Cardiomegaly, Edema, and Pleural Effusion, the mean sensitivities of the models of 0.750, 0.617, 0.712, and 0.806 respectively, are higher than the mean radiologist sensitivities of 0.646, 0.485, 0.710, and 0.761 (at the mean radiologist specificities of 0.806, 0.924, 0.925, and 0.883 respectively). On Consolidation, the mean sensitivity of the models of 0.443 is lower than the mean radiologist sensitivity of 0.456 (at the mean radiologist specificity of 0.935).

\subsection*{Analysis}
Because our primary performance measures do not reveal any information on patterns of mistakes or systematic biases, we qualitatively analyzed chest x-rays where the model output was wrong compared to ground truth diagnosis of consolidation. We used CAMs to reason about model mistakes. The analysis revealed that the type of model mistakes could be pooled into four distinct categories as shown in Table \ref{tbl:cam_analysis}. Each chest x-ray was categorized into one or more of four categories: Failure to correctly localize the consolidation, Failure to confidently detect consolidation, Mistaking a mimicking feature for consolidation, Mistaking a non-mimicking feature for consolidation. The most common mistake was failure to detect to consolidation, and as can be expected this was often the case for faint or small consolidations.

 
\subsection*{Discussion}
Given the variety of healthcare systems and patient populations, it is critical for deep learning models in healthcare to be able to generalize to new patient populations from different institutions \cite{kelly, mlnotdl}. There have been several studies investigating the generalization of models to different institutions. Particularly for chest x-ray interpretation models, \citet{zech_variable_2018} trained image classifiers on chest x-ray from three different institutions and found that  models trained on data from one institution failed to generalize to other institutions. \citet{mlnotdl} raised concerns about whether deep learning based approaches could generalize to smaller healthcare institutions with limited data. \citet{kelly} detailed limitations of deep learning towards generalization to new populations given that the models may learn confounders present in one population. However, \citet{breastcancer} recently showed that the performance of deep learning models on the task of breast cancer detection entirely trained on data from the UK generalized to healthcare data from the US. \citet{kim_jang_kim_shin_park_2019} reported that only 6\% of studies evaluating the performance of AI algorithms for diagnostic analysis of medical images performed external validation.

In our study, we found that CheXpert-trained models demonstrated generalizability to another institution's data without any additional site specific training. Furthermore, the models exceeded radiologists on sensitivity for majority of the tasks when thresholded on radiologists' specificity despite not having been trained on the dataset. The CAMs demonstrate that the model is learning clinically relevant information in the chest x-rays and not confounders.

\section{Limitations}
Our primary assumption in testing the generalization of these models for these different tasks and circumstances is that these models had not been exposed to data used for the external test sets. All models used in the study were trained exclusively to classify CheXpert pathologies (and did not include TB or NIH-specific pathologies): we verified that the output of all models had complete intersection with the CheXpert pathologies.

Furthermore, the results of our study do not suggest guaranteed generalization of chest x-ray models to new clinical settings; future work should evaluate evaluate the performance in clinical trials for further verification, a necessary step for the successful translation of diagnostic or predictive artificial intelligence tools into practice \citep{park_methodologic_2018}.

\section{Conclusion}
Despite advances in the performance of chest x-ray algorithms \citep{lakhani_deep_2017, kallianos_how_2019, shih_augmenting_2019, kashyap_artificial_2019, qin_using_2019, qin_computer-aided_2018}, the ability of these models to generalize has not been systematically explored. The purpose of this study was to systematically evaluate the generalization capabilities of existing models to (1) detect diseases not explicitly included in model development, (2) smartphone photos of x-rays, and (3) x-rays from institutions not included in model development. Our results suggest the possibility for existing chest x-ray models to generalize to new clinical settings without fine-tuning.

Deep learning models, including for chest x-ray interpretation, have been criticized for their inability to generalize to new clinical settings \cite{kelly}. For instance, \citet{zech_variable_2018} reported that chest x-ray models failed to generalize to new populations or institutions separate from the training data, relying on institution specific and/or confounding cues to infer the label of interest. In contrast, our results suggest that existing models may generalize across institutions, modalities, and diseases without further engineering. Importantly, in evaluation of the models there was no indication of bias toward institution specific features in model decision making or a reliance on unrelated features for classification as evident from the class activation maps.

Our systematic examination of the generalization capabilities of existing models can be extended to other tasks in medical AI \citep{rajpurkar2018mura, hannun2019cardiologist, park2019deep, uyumazturk2019deep, varma2019automated, duan2019clinical, topol2019high}, and provide a framework for tracking technical readiness towards clinical translation.

\begin{acks}
We would like to acknowledge the Stanford Machine Learning Group (stanfordmlgroup.github.io) and the Stanford Program for Artificial Intelligence in Medicine and Imaging for infrastructure support (AIMI.stanford.edu).

We would also like to acknowledge those among the top submitters in the competition who helped us understand the data and techniques used in their models: Wenwu Ye from JF healthcare, Hieu Pham from the Medical Imaging Team at Vingroup Big Data Institute (VinBDI), Desmond from Beihang University, Vu Hoang and Hoang Ngoc Nguyen from the VinBrain Applied Scientist Team.
\end{acks}

\bibliographystyle{ACM-Reference-Format}
\bibliography{refs}


\end{document}